\documentclass[letterpaper, 10 pt, conference]{ieeeconf}  % Comment this line out if you need a4paper

\usepackage{graphicx}
\usepackage[bottom]{footmisc}
\usepackage{subfigure}
\usepackage{balance}
\IEEEoverridecommandlockouts                              % This command is only needed if 
\title{\LARGE \bf
EEG Study of the Influence of Imagined Temperature Sensations on Neuronal Activity in the Sensorimotor Cortex
}

\author{Anton Belichenko$^{1,2}$ Daria Trinitatova$^{1}$, Aigul Nasibullina$^{3}$, Lev Yakovlev$^{3}$, and Dzmitry Tsetserukou$^{1}$
\thanks{$^{1}$The authors are with the Intelligent Space Robotics Laboratory, Skolkovo Institute of Science and Technology (Skoltech), 121205 Moscow, Russia. 
        {\tt\small daria.trinitatova, d.tsetserukou$\}$@skoltech.ru}}%
\thanks{$^{2}$Anton Belichenko is with Pirogov Russian National Research Medical University, 117513 Moscow, Russia.
        {\tt\small a.belichenko1507@gmail.com}}%
\thanks{$^{3}$The authors are with Vladimir Zelman Center for Neurobiology and Brain Rehabilitation, Skolkovo Institute of Science and Technology, Moscow, Russia.
       }%
}

\begin{document}

\maketitle
\thispagestyle{empty}
\pagestyle{empty}

%%%%%%%%%%%%%%%%%%%%%%%%%%%%%%%%%%%%%%%%%%%%%%%%%%%%%%%%%%%%%%%%%%%%%%%%%%%%%%%%
\begin{abstract}

Understanding the neural correlates of sensory imagery is crucial for advancing cognitive neuroscience and developing novel Brain-Computer Interface (BCI) paradigms. This study investigated the influence of imagined temperature sensations (ITS) on neural activity within the sensorimotor cortex. The experimental study involved the evaluation of neural activity using electroencephalography (EEG) during both real thermal stimulation (TS: 40 °C Hot, 20 °C Cold) applied to the participants' hand, and the mental temperature imagination (ITS) of the corresponding hot and cold sensations. The analysis focused on quantifying the event-related desynchronization (ERD) of the sensorimotor $\mu$-rhythm (8-13 Hz). The experimental results revealed a characteristic $\mu$-ERD localized over central scalp regions (e.g., C3) during both TS and ITS conditions. Although the magnitude of $\mu$-ERD during ITS was slightly lower than during TS, this difference was not statistically significant ($p>.05$). However, ERD during both ITS and TS was statistically significantly different from the resting baseline ($p<.001$). These findings demonstrate that imagining temperature sensations engages sensorimotor cortical mechanisms in a manner comparable to actual thermal perception. This insight expands our understanding of the neurophysiological basis of sensory imagery and suggests the potential utility of ITS for non-motor BCI control and neurorehabilitation technologies.

\end{abstract}

\section{INTRODUCTION}

Understanding the neural mechanisms of sensory processing and mental imagery is crucial for advancing cognitive science and developing sophisticated human-machine systems. Mental imagery, the ability to form internal representations of physical activities such as actions, touches, or sounds, is a crucial cognitive process. This cognitive ability is related to a wide range of functions, including memory, creativity, motor control, navigation, arithmetic and moral decision-making \cite{pearson2019human}. While the neural underpinnings of motor and visual imagery are comparatively well understood, the somatosensory domain remains less explored. The sensorimotor cortex, traditionally associated with motor control and tactile processing, is key to these functions. It is well-established that imagining motor actions (Motor Imagery) or tactile sensations (Tactile Imagery) modulates activity in this area \cite{munzert2009cognitive,ladda2021using,yakovlev2023event}. 
\begin{figure}[!t]
    \centering
    \includegraphics[width=0.9\linewidth]{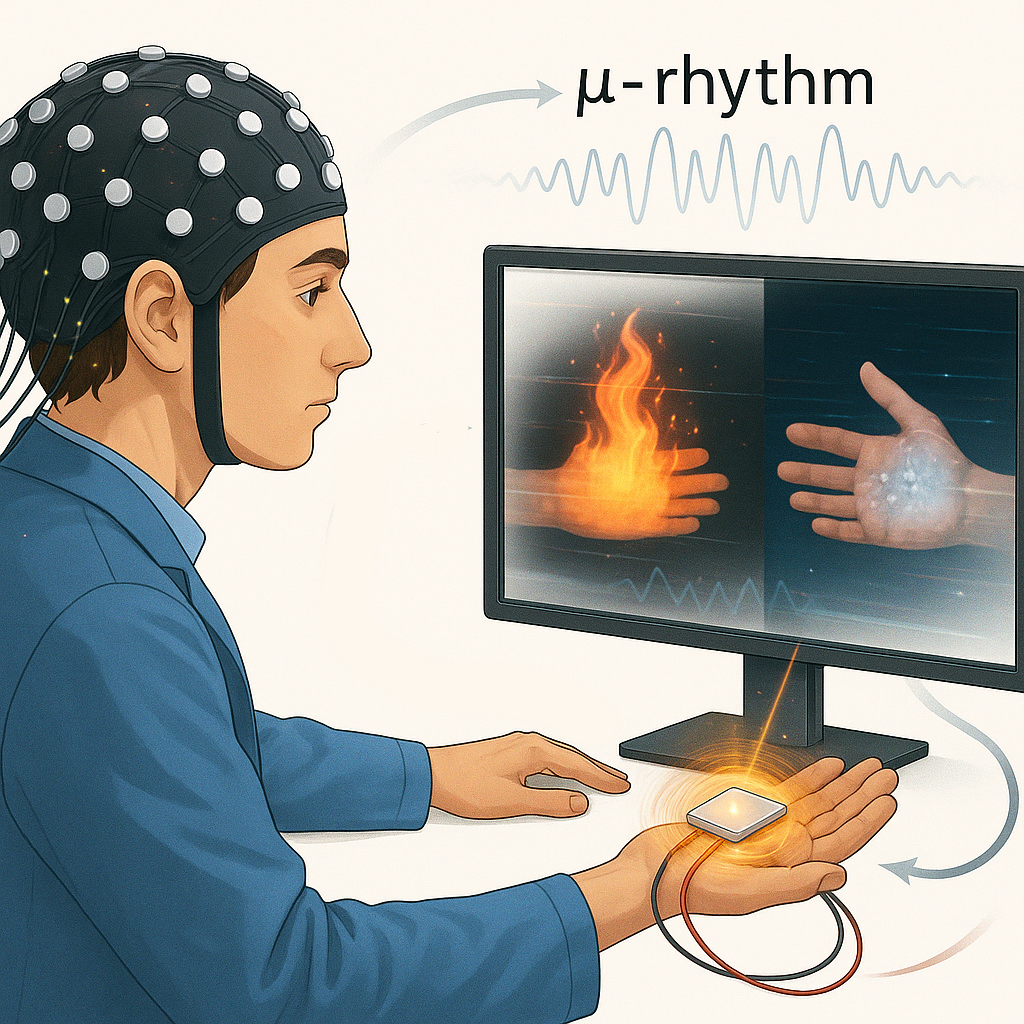}
    \caption{The concept of EEG study for detecting Event-Related Desynchronization in the $\mu$ rhythm induced by temperature stimulation.}
    \label{fig:concept}
\end{figure}

Motor imagery (MI), a widely studied paradigm, is known to induce cortical neuroplasticity and facilitate motor learning \cite{ladda2021using,ruffino2017neural}. Depending on the primary sensory modality, MI generally distinguished into kinesthetic and visual types \cite{stevens2005interference}. Kinesthetic MI, in particular, engages motor circuits that overlap with those for voluntary movement. Since this process involves imagining muscle and skin sensations, it plausibly engages somatosensory areas,  which is supported by recent findings that somatosensory signals contribute to cortical excitability during MI \cite{kaneko2014motor,yakovlev2019corticospinal}.

In contrast to MI, tactile imagery (TI) has has received less scientific focus. Neuroimaging investigations of TI have predominantly utilized fMRI methods \cite{yoo2003neural,schmidt2019somatotopy}, with fewer systematic studies employing EEG. Nonetheless, these studies have shown that TI activates the sensorimotor cortex and alters coupling between cortical areas \cite{yoo2003neural,schmidt2019somatotopy,bashford2021neurophysiological}. This form of imagery, also termed somatosensory attention orientation, shows promise as an alternative BCI modality to MI \cite{yao2017stimulus,yao2022performance}. These findings suggest a similarity in neural patterns between actual sensory perception and TI tasks, mirroring MI studies in which overlapping brain regions are active during both physical movements and motor imagery.

Modulation within the sensorimotor cortex frequently appears as a desynchronization of the $\mu$-rhythm (8–13 Hz) in electroencephalography (EEG) signals, termed Event-Related Desynchronization (ERD). This $\mu$-ERD, widely interpreted to reflect increased processing within the sensorimotor network, has become a cornerstone for the development of BCIs based on motor imagery \cite{singh2021comprehensive,al2021systematic}. A characteristic marker of sensorimotor processing is a transient decrease in EEG/magnetoencephalography power within a specific frequency band. This ERD, along with its counterpart, event-related synchronization (ERS), signifies changes in the correlated activity of the underlying neuronal populations, reflecting increased and decreased cortical processing, respectively \cite{pfurtscheller1999event}. Sensorimotor rhythms, oscillatory activities typically in the $\alpha$ (8–12 Hz, often called $\mu$-rhythm) and $\beta$ (13–30 Hz) ranges, are observed in cortical sensorimotor areas during various states, including voluntary movement, motor imagery, movement observation, and tactile stimulation \cite{hari1997human}.

However, the role of the sensorimotor cortex in imagining other sensor modalities, particularly temperature, remains largely unexplored. Although thermal sensation is a fundamental sensory experience, its precise neural representation and especially the neural correlations of imagined temperature sensations (ITS) is not yet fully understood \cite{watanabe2025spatial}. This leads to key questions: Does the brain simulate temperature imagery using mechanisms similar to those involved in motor imagery or tactile imagery?  Specifically, does imagining temperature modulate the $\mu$-rhythm in a manner comparable to actual thermal stimulation or other imagery types?

Investigating the neural correlations of ITS can provide critical insights into the cortical representation of diverse sensory experiences and the functional plasticity of the sensorimotor system. Understanding how purely sensory imagery, such as imagining heat or cold, affects brain activity could open new avenues for BCI applications. If distinct and reliable neural patterns, such as $\mu$-ERD, can be identified and decoded from single-trial EEG data during ITS, this could offer alternative control strategies for individuals unable to perform motor imagery. In addition, it could be used in neurofeedback applications aimed at modulating sensorimotor cortex excitability for therapeutic purposes, aligning with the development of smart systems for enhancing human capabilities.

This study aims to address these gaps by investigating the impact of imagining distinct temperature sensations (hot and cold) on sensorimotor $\mu$-rhythm activity recorded via multi-channel EEG. We compare the ERD patterns evoked during ITS with those observed during actual thermal stimulation (TS) the participant's hand. Building on findings suggesting that imagined and real tactile stimuli might engage overlapping neural mechanisms (\cite{yakovlev2023event,morozova2024tactile}), we explore whether a similar principle applies to the thermal domain. This study could contribute to a broader understanding of the neural underpinnings of sensory imagination, especially in the understudied thermal modality, and facilitate the design of future BCI and neurorehabilitation technologies in which motor imagination is already actively used as a training tool \cite{mulder2007motor,almufareh2023leveraging}.

\section{System Architecture}
For this study, an experimental setup was used to analyze EEG rhythms of the participants during thermotactile stimulation and imagination of temperature sensations without physical and visual stimulation.
\begin{figure}[!h]
    \centering
    \includegraphics[width=0.95\linewidth]{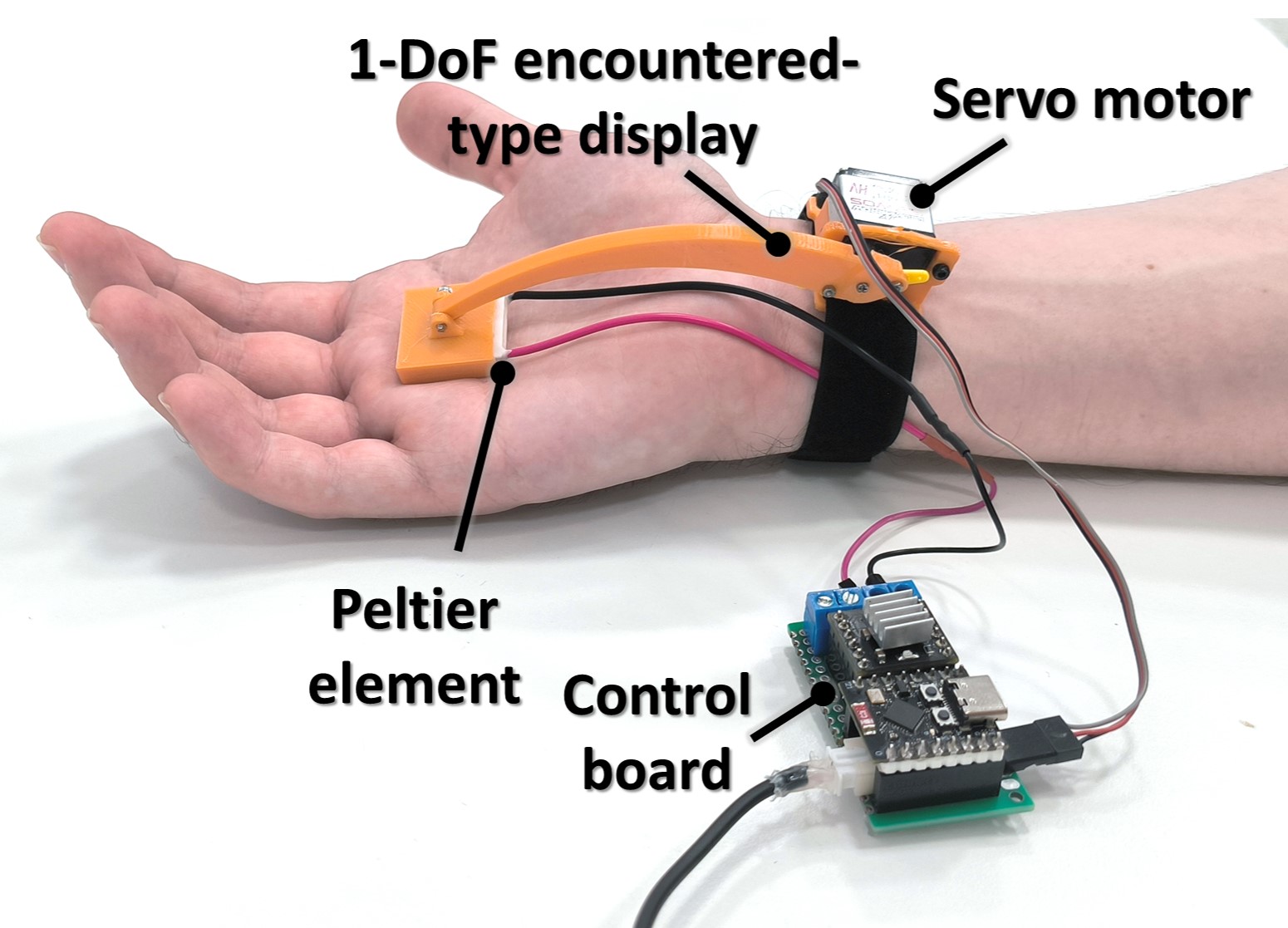}
    \caption{A designed 1-DoF encountered-type haptic display for providing thermal feedback.}
    \label{fig:haptic_display}
\end{figure}
\subsection{Tactile Display for Temperature Stimulation}\label{tactile display}
For providing thermotactile stimuli for short-term use, we utilized a Peltier module (TEC1-03108, 20×20 mm) both for cold and warm sensations. The application of Peltier modules has been successfully employed in previous studies of thermal perception \cite{davis1998functional,mulders2020dynamics,watanabe2025spatial}.

To eliminate the residual effect of temperature stimulation on the skin during constant interaction of the Peltier element with the palm of the hand, we designed an encountered-type device for the palm with 1 DoF that holds the Peltier element on the end effector (Fig. \ref{fig:haptic_display}). The mechanism is actuated by the Blue Arrow D0474HT-HV micro-servo motor and provides non-contact and contact states between the Peltier element and the user's skin. The designed tactile display is attached to the user wrist using elastic tapes. The Peltier element is driven with pulse width modulation (PWM) control using an ESP32 microcontroller with a DC motor driver (DRV8833). This configuration enabled smooth temperature adjustments and reversal of current polarity, necessary to rapidly switch between heating and cooling modes.
For our study, the Peltier element was operated in cooling (20 °C) and heating mode (40 °C). Temperature measurements were made using a thermal imaging camera. 

\begin{figure*}
    \centering
    \includegraphics[width=0.98\linewidth]{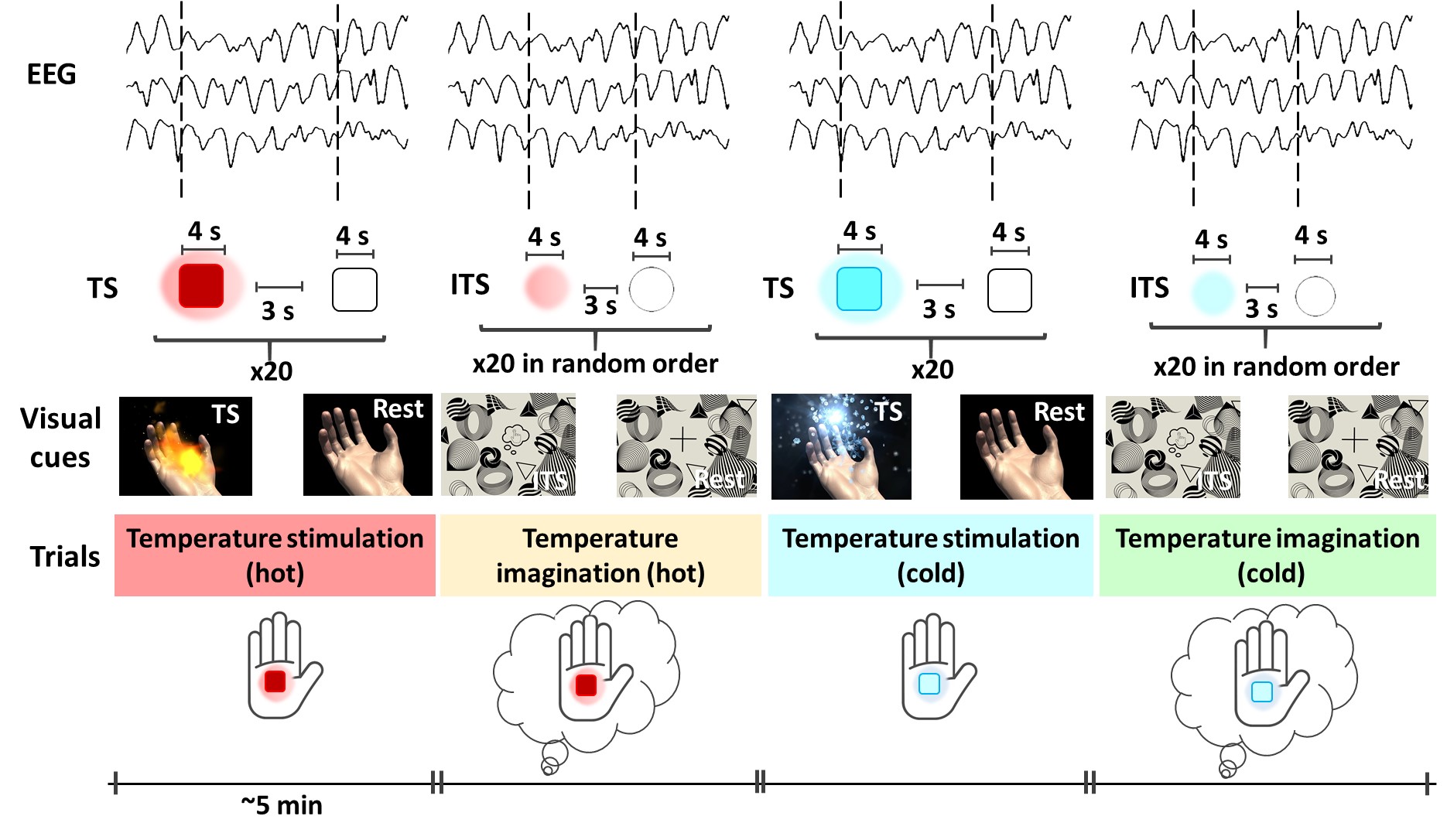}
    \caption{A scheme of the experimental session. The session consisted of four consecutive conditions: hot temperature stimulation (TS hot), imagined hot temperature stimulation (ITS hot), cold temperature stimulation (TS cold), imagined cold temperature stimulation (ITS cold). Each condition included a mixture of 20 somatosensory trials (TS or ITS) and 20 rest trials.}
    \label{fig:exp}
\end{figure*}
\subsection{EEG Data Acquisition and Preprocessing}

Continuous EEG data were obtained using a 48-channel NVX-52 DC amplifier (MCS, Russia). Passive Ag/AgCl electrodes were placed according to the International 10-10 system, with electrode impedance maintained below 15 $k\Omega$ throughout the experiment to ensure high-quality signals. Data were sampled at 500 Hz. Hardware filters applied during acquisition included a 0.1 Hz high-pass and a 100 Hz low-pass filter. Linked mastoids (TP9/TP10) served as the reference electrodes during recording.

Offline preprocessing was performed using the MNE-Python library \footnote{https://github.com/mne-tools/mne-python} with the following steps:
\begin{itemize}
    \item Filtering: Raw data were band-pass filtered between 1 Hz and 40 Hz using a Finite Impulse Response (FIR) filter. A notch filter at 50 Hz was applied to remove power line interference.
    \item Artifact removal: Independent component analysis (ICA) was used to identify and remove stereotyped artifacts, such as eye blinks and horizontal eye movements. Visual inspection was also used to reject epochs contaminated by excessive muscle activity or other non-stereotyped noise. Noisy or malfunctioning channels were interpolated using spherical spline interpolation.
    \item Re-referencing: Data were re-referenced to the common average reference (CAR).
    \item Epoching: The continuous preprocessed data were segmented into epochs time-locked to the onset of each condition's stimulus (TS, ITS, Rest). Each epoch spanned from -1 $s$ pre-stimulus (baseline) to +4 $s$ post-stimulus, covering the task duration.
\end{itemize}

\section{User Study}

\subsection{Participants}

Fifteen healthy volunteers (9 females, aged from 20 to 35 years) participated in the study. All participants were right-handed and reported no history of neurological or psychiatric disorders. The participants were naive about the specific experimental task involving temperature imagery. Before the experiment, participants provided their written informed consent.

\subsection{Experimental Conditions}

The experiment used a within-subject design comparing brain activity across three main conditions: real temperature stimulation (TS), imagined temperature sensation (ITS), and a resting baseline condition (Control).

\subsubsection{Real Temperature Stimulation (TS)} 
Controlled thermal stimuli were delivered to the palm of the participant’s dominant hand using the designed encountered-type thermo-tactile display (Section \ref{tactile display}). For each trial, the end effector first reached a contact position with the palm and a TS was provided for 4 $s$, and then returned to the non-contact position. Two distinct temperature conditions were used: Hot (target temperature: 40 °C) and Cold (target temperature: 20 °C). These temperatures were selected based on pilot testing to provide clear thermal sensations without causing discomfort or pain. Each stimulation period lasted 4 $s$. During TS, participants received the corresponding visual feedback on the screen: a virtual hand representation with dynamic flame effects for hot TS or ice effects for cold TS, implemented in Unity 3D.

\subsubsection{Imagined Temperature Sensation (ITS)} 
Participants were instructed to vividly imagine the sensation of warmth (corresponding to the 40 °C stimulus) or coldness (corresponding to the 20 °C stimulus) on their dominant hand, in the absence of any physical stimulation.  Each imagery period also lasted 4 $s$. During ITS trials, participants viewed a neutral abstract visual pattern on the screen to minimize the external visual influence related to temperature.
\subsubsection{Control Condition} 
Participants were instructed to relax while focusing on the center-cross image (in case of ITS) or a neutral image of a hand (in case of TS) displayed on the screen. Each Control trial lasted 4 $s$ and served as a baseline reference state for subsequent ERD/S calculations.

\subsection{Experimental Procedure}

The participants were comfortably seated in an armchair in a quiet laboratory environment, facing a computer screen positioned approximately 1 $m$ away. After the electrode setup and initial instructions, they completed a brief familiarization phase to familiarize themselves with both the real thermal stimuli and the requirements for the mental imagery task.

\begin{figure}[!t]
  \centering
  \subfigure[]{\includegraphics[width=0.625\linewidth]{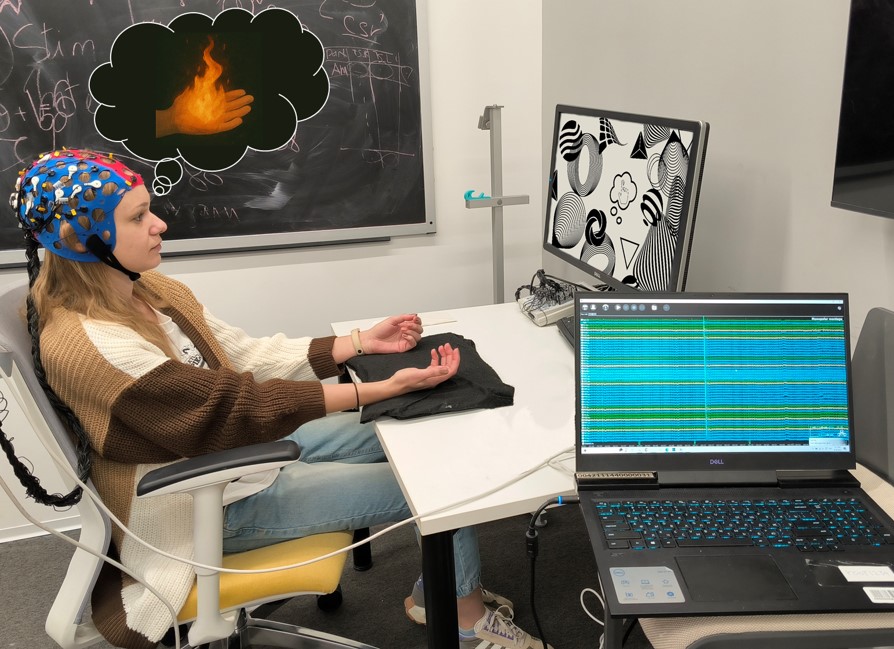}}
   \subfigure[]{\includegraphics[width=0.335\linewidth]{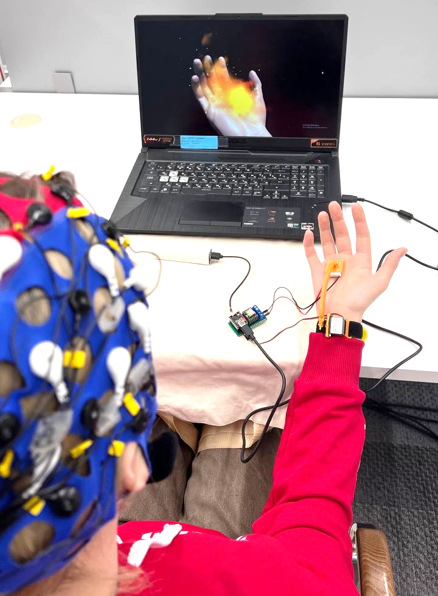}}
  \caption{Experimental setup during the EEG study. A participant receives either thermal stimulation using the designed tactile display (b) or imagines temperature sensations during EEG recordings (a).}\label{fig:us}
\end{figure}

The experimental session consisted of four stimulation blocks: each focusing on one condition pair: TS-Hot/Control, ITS-Hot/Control, TS-Cold/Control, and ITS-Cold/Control (Fig. \ref{fig:exp}).  Within each block, 20 trials of the active condition (TS or ITS, lasting 4 $s$) were randomly interleaved with 20 trials of the Control condition (lasting 4 $s$). A 3-second inter-trial interval with a fixation cue separated each trial (active or rest). Participants were instructed to remain still, minimize eye movements, maintain fixation when required, and focus their attention fully on physical sensation (during TS) or vividly imagined sensation (during ITS) throughout the duration of a 4-second active trial (Fig. \ref{fig:us}). In total, each stimulation session lasted about 5 minutes. %Subjective vividness ratings for imagery were collected after completing each ITS block.

\subsection{Data Analysis: ERD/S Quantification}

The primary analysis focused on quantifying Event-Related Desynchronization (ERD) and Synchronization (ERS) in the $\mu$ frequency band (8-13 Hz). Individual $\mu$-rhythm frequency ranges were determined for each participant based on visual inspection of their power spectra in the resting state.
\begin{itemize}
    \item Spectral Power Estimation: For each epoch and channel, the time-varying spectral power within the individually defined $\mu$ band was calculated using the Morlet wavelet transform.
    \item ERD/S Calculation: The ERD/S was computed as the percentage power change during the activation period (defined as 0.5–4 $s$ post-stimulus onset) relative to the power during a pre-stimulus baseline interval (-1 to 0 seconds) using equation:
\begin{equation}
ERD/S = \frac{(PSD_{act} - PSD_{rst}) }{PSD_{rst}} \cdot 100,   
\end{equation}
where PSD$_{act}$ is the across-epoch average power for the TS, ITS trials for each channel, and PSD$_{rst}$ defines the average spectral power calculated for trials corresponding to the resting state.
    \item Statistical analysis: %To identify statistically significant ERD/S effects across time and scalp locations, and to compare mean ERD/S magnitudes between conditions within specific sensorimotor channels of interest (ITS vs. Control, TS vs. Control, ITS vs. TS), the nonparametric Wilcoxon signed-rank test was utilized with a significance level set at $\alpha<.05$. %Nonparametric cluster-based permutation tests were used to correct for multiple comparisons across channels and time points. 
    For comparing mean ERD/S magnitudes between conditions (ITS vs. Control, TS vs. Control, ITS vs. TS) within specific sensorimotor channels of interest (e.g., C3), the Wilcoxon signed-rank test was utilized. The significance level was set at $\alpha<.05$.
    \item Topographic mapping: To visualize the spatial distribution of sensorimotor modulation, scalp topograms illustrating mean $\mu$-ERD/S values in the corresponding time windows were plotted for each experimental condition.

\end{itemize}

\section{Experimental Results}

\subsection{$\mu$-Rhythm Desynchronization during ITS and TS}

Consistent with our hypothesis, both the real temperature stimulation (TS) conditions (Hot 40 °C and Cold 20 °C) and the imagined temperature sensation (ITS) conditions elicited a clear Event-Related Desynchronization (ERD) in the $\mu$ frequency band (8-13 Hz). This desynchronization, reflecting a decrease in $\mu$-rhythm power during the task compared to the resting baseline, persisted throughout the entire 4-second trial interval. This sustained effect is similar to the ERD patterns observed during sustained tactile imagery tasks (\cite{yakovlev2023event,morozova2024tactile}). 

\subsection{Topographic Distribution of $\mu$-ERD/S}

\begin{figure}
    \centering
    \includegraphics[width=0.98\linewidth]{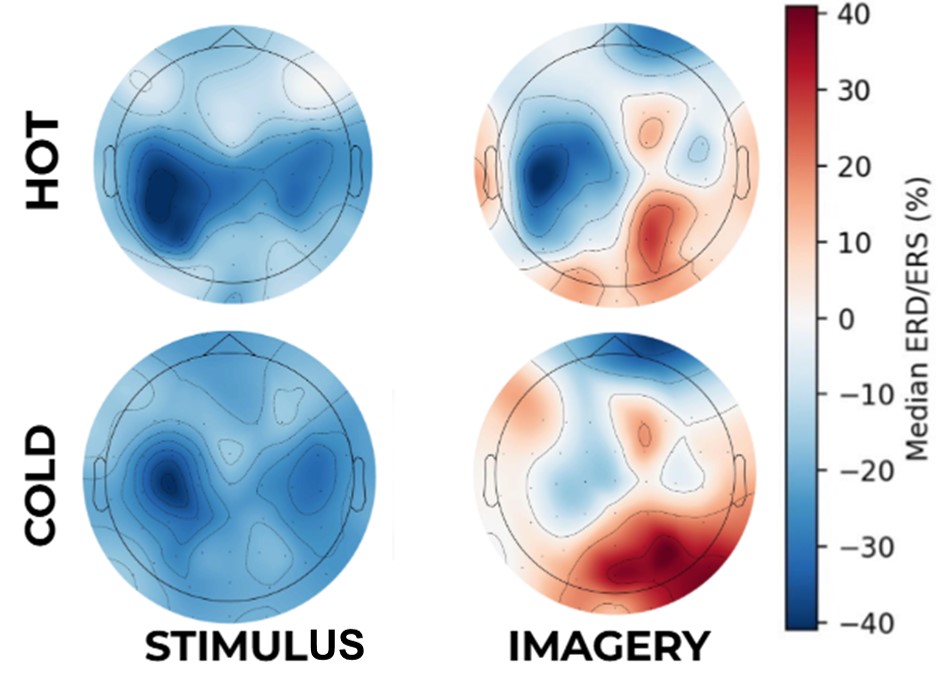}
    \caption{Median event related desynchronization/synchronization (ERD/S) topomaps of sensorimotor EEG $\mu$-rhythmic activity in TS and ITS conditions for all subjects. Blue corresponds to the level of desynchronization (ERD), red corresponds to
    synchronization (ERS).}
    \label{fig:Topographic maps}
\end{figure}

The observed neural activation ($\mu$-ERD) was topographically focused on the contralateral sensorimotor cortex, with desynchronization patterns evident across a cluster of electrodes, including C1, C3, C5, CP3, and CP5. The strongest desynchronization was typically observed around the contralateral central electrodes, particularly C3 (assuming the dominant right hand was the focus of stimulation and imagery), which is consistent with the findings of studies on real and imagined tactile stimuli (\cite{yakovlev2023event,morozova2024tactile}). In contrast, analysis of the ipsilateral electrode C4 and the midline electrode Cz did not reveal statistically significant ERD patterns ($p>.05$). This topographical specificity suggests that the observed neural modulation is well localized to the core sensorimotor area responsible for hand sensation, rather than being a generalized or diffuse brain response. Figure \ref{fig:Topographic maps} illustrates the average scalp topographies of $\mu$-ERD for TS and ITS conditions, showing separate maps for Hot and Cold. The spatial patterns of activation appeared to be qualitatively similar between real stimulation (TS) and imaginary sensation (ITS) for both temperatures, suggesting the involvement of overlapping cortical substrates during both types of thermal perception.

\subsection{Statistical Comparisons of $\mu$-ERD Intensity}

\begin{figure}
    \centering
    \includegraphics[width=0.98\linewidth]{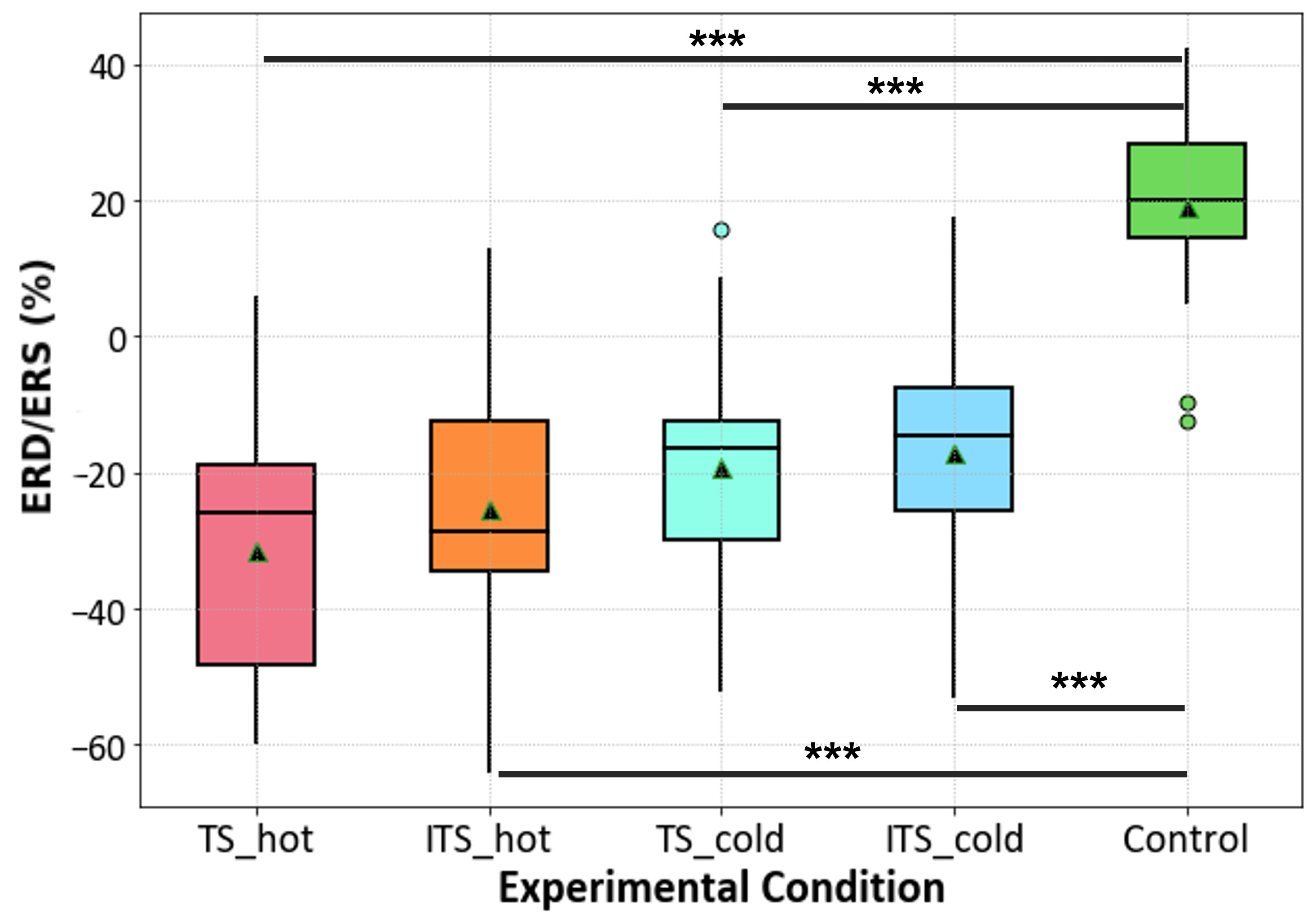}
    \caption{Averaged ERD-values in the C3 channel for explored conditions (Hot temperature stimulation (TS hot), Cold temperature stimulation (TS cold), Hot imagined temperature sensations (ITS hot), Cold imagined temperature sensations (ITS cold), and Resting state (Control). Black triangles represent mean values. Asterisks display the statistical significance of the results ($^{***}:p \leq .001$).}
    \label{fig:bp}
\end{figure}

The clear topographic focus, with a robust peak at C3, guided our decision to select this channel for the primary quantitative analysis comparing the experimental conditions. The distribution of the ERD values obtained in the C3 channel for the conditions explored is shown in Fig. \ref{fig:bp}. According to the Wilcoxon signed-rank test, there were significant differences between the experimental conditions in the magnitude of $\mu$-ERD in the C3 channel ($\chi^2$=29.067, $p<.001$). The pairwise comparison revealed that the magnitude of $\mu$-ERD during both TS and ITS conditions was significantly stronger compared to the resting baseline condition (Table \ref{statistics}). This indicates that both real perception and mental imagination of temperature induced a reliable modulation of sensorimotor cortical activity.
\begin{table}[!h]
\caption{Statistical Analysis of Sensorimotor ERD over C3 Channel }
\label{statistics}
\centering
\begin{tabular}{ccc}
\hline
\textbf{Comparisons} & \textbf{Wilcoxon signed-rank test
} & \textbf{$p$-value}\\
\hline
$TS_{hot}$ vs $Control$ & 0.0 & $<.001$\\
$TS_{cold}$ vs $Control$ & 4.0 & $<.001$\\
$ITS_{hot}$ vs $Control$ & 2.0 & $<.001$\\
$ITS_{cold}$ vs $Control$ &0.0& $<.001$\\
$TS_{hot}$ vs $ITS_{hot}$ &45.0 & .42\\
$TS_{cold}$ vs $ITS_{cold}$ & 56.0& .85\\
\hline
\end{tabular}\label{tab:stat}
\end{table}

When directly comparing the intensity of $\mu$-ERD between the real TS and ITS conditions, the desynchronization during ITS tended to be slightly less pronounced than during TS for both Hot and Cold scenarios (Fig. \ref{fig:bp}). This trend is consistent with findings comparing real and imagined tactile stimulation \cite{yakovlev2023event}. However, in our study, the difference in ERD magnitude between TS and ITS was statistically insignificant ($p>.05$, based on Wilcoxon signed-rank test).

\subsection{Limitations}
This study analyzed data from 15 participants. Although the sample size is modest, highly statistically significant effects in $\mu$-ERD were observed (see Table \ref{tab:stat}), which confirms the robustness of the core phenomenon. However, we recognize that a larger sample of participants would be necessary to explore the full range of individual differences and enhance generalizability of the findings obtained. Similarly, our focus on a homogeneous cohort of young adults (20–35 years) was a deliberate choice to minimize confounding variables, meaning that age-related effects were not the subject of this particular investigation. Future work should aim to replicate these findings in larger and more age-diverse populations.

\section{Conclusions}

The results presented in this study demonstrate that ITS reliably induces a significant desynchronization of the $\mu$-rhythm (ERD) in the human EEG. The peak of this activation was consistently and most significantly identified at the C3 electrode, which aligns with the cortical representation of the stimulated right hand. In contrast, the ipsilateral hemisphere (e.g., C4) showed no significant modulation. This highly specific pattern reinforces the notion that sensory imagery engages the same core cortical mechanisms responsible for actual perception. This principle is well-established for motor imagery (\cite{munzert2009cognitive,ladda2021using}) and has recently been extended to tactile imagery (\cite{yakovlev2023event,morozova2024tactile}). The functional specificity of this response, confined to the contralateral sensorimotor hand area, suggests that ITS engages the brain in a precise, rather than diffuse, manner.

A key novel contribution of this work lies in the observed asymmetry between heat and cold imagery, namely, heat-related ITS produced more robust ERD than cold. This asymmetry is consistent with established theories in evolutionary neuroscience and psychophysiology. From an adaptive perspective, the increased cortical response to imagined heat may reflect an evolved prioritization of hyperthermia detection, due to its immediate physiological risks, such as dehydration, cardiovascular strain, and systemic thermal stress \cite{crandall2015human}. This prioritization may result in more vivid and accessible heat-related engrams during imagery, as reflected in the greater $\mu$-ERD.

Our findings indicate that the neural mechanisms involved during the imagination of temperature share considerable overlap with those involved in processing real thermal stimulation, suggesting that sensory imagination reactivates stored representations of bodily experience. This interpretation is consistent with the thermoregulation perception models\cite{havenith2001individualized}, which propose a tight coupling between subjective thermal experience and physiological states. As with motor imagery, where imagined actions recruit execution-related pathways, thermal imagery appears to simulate somatosensory and interoceptive responses associated with heat and cold.

These findings could have important implications for the fields of BCI and neurorehabilitation. The reliable involvement of the sensorimotor cortex through ITS suggests its dual potential in neurotechnology: first, as an intuitive control signal for BCI systems, particularly for users with motor impairments, and second, as a novel modality for neurofeedback and rehabilitative protocols designed to modulate cortical plasticity.

Our future work will focus on characterizing the ITS-induced neural response in more detail. Exploring the ability to discriminate between imagined hot and cold sensations based on EEG patterns using machine learning techniques could reveal whether different thermal qualities possess distinct neural signatures, complementing studies focused on discriminating real thermal stimuli. Exploring the integration of ITS into functional BCI prototypes represents a promising direction to translate these findings into practical applications. 

\section{Acknowledgements} 
Research reported in this publication was financially supported by the RSF grant No. 24-41-02039.

\addtolength{\textheight}{-12cm}   % This command serves to balance the column lengths

%%%%%%%%%%%%%%%%%%%%%%%%%%%%%%%%%%%%%%%%%%%%%%%%%%%%%%%%%%%%%%%%%%%%%%%%%%%%%%%%

%%%%%%%%%%%%%%%%%%%%%%%%%%%%%%%%%%%%%%%%%%%%%%%%%%%%%%%%%%%%%%%%%%%%%%%%%%%%%%%%

%%%%%%%%%%%%%%%%%%%%%%%%%%%%%%%%%%%%%%%%%%%%%%%%%%%%%%%%%%%%%%%%%%%%%%%%%%%%%%%%

\bibliographystyle{IEEEtran}
\balance
\bibliography{ref}

\end{document}